\documentclass[aps,floats,epsf,twocolumn]{revtex4}
 \usepackage{graphics}

\begin{document}

\title{Zero modes and charged Skyrmions in graphene bilayer}
\author{Chi-Ken Lu$^1$ and Igor F. Herbut$^{1,2}$}

\affiliation{$^1$ Department of Physics, Simon Fraser University,
 Burnaby, British Columbia, Canada V5A 1S6 \\ $^2$ Kavli Institute for
 Theoretical Physics, University of California, Santa Barbara, CA
 93106, USA}

\begin{abstract}
 We show that the electric charge of the Skyrmion in the vector order parameters that characterize the quantum anomalous spin Hall state and the layer-antiferromagnet in a graphene bilayer is four and zero, respectively. The result is based on the demonstration that a vortex configuration in two broken symmetry states in bilayer graphene with the quadratic band crossing has the number of zero modes doubled relative to the single layer. The doubling can be understood as a result of Kramers' theorem implied by the ``pseudo time reversal" symmetry of the vortex Hamiltonian. Disordering the quantum anomalous spin Hall state by Skyrmion condensation should produce a superconductor of an elementary charge $4e$.
\end{abstract}
\maketitle

\vspace{10pt}

A Bernal-stacked graphene bilayer is an interesting example of a two-dimensional system of electrons
with the quadratic band crossing.  When the Fermi energy is tuned to the
band crossing point, a simple argument shows that the non-interacting ground state is unstable,
and the gap in the spectrum opens at an infinitesimal repulsive
interaction.\cite{sun}  Indeed, there have been several recent experiments that find a broken-symmetry insulating ground state and the concomitant spectral gap in graphene bilayers.\cite{velasco, feldman, weitz, freitag, mayorov} The exact nature of the insulating ground state is presently a point of contention, since, at least for the Coulomb repulsion between electrons, the ground states with different broken symmetries may lay rather close in energy.\cite{nandkishore, zhang, jung, zhang1, lemonik, vafekyang, vafek, uebelacker}

Since the different broken symmetry states in bilayer graphene appear to be nearly degenerate, their topologically nontrivial spatial configurations,  such as vortices, merons, and Skyrmions, which would represent stable thermal or quantum fluctuations, become highly relevant. These topological defects in single layer of graphene are known to carry non-trivial quantum numbers,\cite{hou, ryu, herbut1} since the corresponding Dirac Hamiltonian is linear in momentum and closely related to the example introduced by Jackiw and Rossi in their classic study of charge fractionalization in two dimensions.\cite{jackiwrossi} In the single layer graphene, however, at least without a magnetic field, any breaking of symmetry requires an interaction strength which appears to be too high.\cite{herbut2} In the bilayer, on the other hand, the Hamiltonian is quadratic in momentum and without the Lorentz symmetry. As a result, its spectrum in the presence of topologically nontrivial configurations of the order parameters (masses) is unknown. In this note we discuss some global features of this spectrum and point out some of its physical consequences.

We first consider the Hamiltonian for the spinless electrons in the bilayer graphene in the presence of a unit vortex in two, out of three, mutually anticommuting mass terms. These masses represent: 1) the state with a broken layer-inversion symmetry (BLIS), and 2) the two valley ferromagnets (VFM$_1$, VFM$_2$) which break both the layer inversion and the time reversal symmetries. The forth possible mass, which commutes with the above three, is the quantum anomalous  Hall (QAH) state. This is the basic Hamiltonian onto which the other problems can, as we show, be mapped. This Hamiltonian, while violating the physical time reversal symmetry, exhibits another hidden antiunitary symmetry, which we name ``pseudo time reversal". Unlike the true time reversal for spinless particles, the pseudo time reversal symmetry squares into $-1$ and implies double degeneracy of the entire spectrum. When combined with the chiral symmetry, this further implies that the zero modes, if present in the spectrum, must come in Kramers  pairs of equal chirality. We obtain the two zero modes analytically for a representative Hamiltonian with the amplitude of the order parameter increasing linearly with the radius. Once found this way, the zero modes cannot shift their energy upon a deformation of the amplitude, as long as the Hamiltonian respects the chiral symmetry. Filling the doubly degenerate zero level fully thus leads to a unit electric charge $e$ being bound to the vortex.

The above result is then used to show that the Skyrmion texture in the vector order parameter that characterizes the quantum anomalous spin Hall (QSH) state in the bilayer graphene carries the electric charge of $4e$. In contrast, the charge of the Skyrmion in the vector order parameter that describes the layered antiferromagnet (LAF) is zero. Quantum disordering the QSH state by the condensation of Skyrmions could thus lead to an exotic non-BCS superconducting state of four-electron composites.\cite{pasha}

{\it Pseudo time reversal symmetry.} -- Assuming that it is  the A-sublattices of the two graphene layers that are coupled by the interlayer hopping $t_\perp$, the low-energy Hamiltonian for the bilayer is
\begin{equation}
H_0 = V_1 \gamma_1 + V_2 \gamma_2,
\end{equation}
with $V_1 = p_1 p_2 /m$, $V_2 = (p_1 ^2 - p_2 ^2 )/ 2m$,  with $p_i$ as the momentum operators measured from the band crossing point, $2m= t_\perp / t^2$, with $t$ as the nearest neighbor intralayer hopping integral.\cite{ed} We chose the representation in which $\gamma_1 = \sigma_3 \otimes \sigma_2$, $\gamma_2 = \sigma_0 \otimes \sigma_1$.\cite{herbut'} $H_0$ acts on the four-component Dirac spinor $\Psi= (b_{1+}, b_{2+}, b_{1-}, b_{2-})^T $, where the first index labels the layer, the second labels the band crossing points $\pm \vec{K}$, and all the components are on the B-sublattices of the two layers.\cite{vafek} We have assumed spinless fermions, and will include the spin shortly. The above Hamiltonian can also be understood in the context of the single layer graphene, as the second-order term in the expansion at the Dirac point in powers of the momenta. Its form is dictated by the $C_{3v}$ symmetry of the honeycomb lattice.

Let us first review some of the familiar symmetries of $H_0$.\cite{herbut'}  In the continuum limit, the translational invariance is an exact (continuous) symmetry. It is generated by the momentum-like operator $\gamma_{35} = \sigma_3 \otimes \sigma_0$. The time reversal, which consists of the exchange of the two Dirac points and the complex conjugation $K$, is also an exact (discrete) symmetry. If we define $\gamma_0= \sigma_0 \otimes \sigma_3$,  $\gamma_3= \sigma_1 \otimes \sigma_2$, and $\gamma_5= \sigma_2 \otimes \sigma_2$, then the time reversal operator may be written explicitly as $I_t = i \gamma_1 \gamma_5 K$, and $\gamma_{35} = -i \gamma_3 \gamma_5$. Note that $I_t ^2 = +1$, since the real spin has not been included yet.  Finally, $\gamma_2$ represents the operation of exchange of the layers, and when accompanied by the inversion of one of the two axis, it is also an exact (discrete) symmetry of $H_0$.

There are only four mass matrices that anticommute with $H_0$ and would gap out the spectrum: $\gamma_0$, $\gamma_3$, $\gamma_5$, and $i\gamma_1 \gamma_2$. They represent, BLIS state, VFM$_1$, VFM$_2$, and the QAH state, respectively, quite similarly to the nomenclature in the single layer graphene.\cite{herbut'}  Only the first and the last state respect translational invariance, and all but the first state violate the time reversal invariance. (See Table 1.) This means that, in contrast to the single layer graphene,\cite{hou} it is impossible to form a four-component bilayer Hamiltonian which would contain a vortex configuration in two of the masses and would simultaneously obey the time reversal symmetry.

We now introduce the notion of the pseudo time reversal invariance. Let us define the antilinear operator
\begin{equation}
A= \gamma_{35} I_t.
\end{equation}
The only mass term that fails to commute with $A$ is the one for QAH state, whereas the other three, together with the Hamiltonian $H_0$, commute with it. (See Table 1.) We therefore choose an arbitrary pair of the first three masses to form a vortex, and study the electronic spectrum in its presence. The vortex Hamiltonian is defined as
\begin{equation}
H_v =  H_0 + \Delta(r) ( \gamma_3 \cos \phi  + \gamma_5 \sin \phi  ),
\end{equation}
where $(r, \phi)$ are the polar coordinates. While $H_v$ now violates both the translational invariance ($\gamma_{35}$) and the time reversal symmetry ($I_t$) separately, it respects  the pseudo time reversal symmetry: $[H_v , A]=0$. Moreover, since the matrix $\gamma_{35}$ can be thought of as the generator of translations, it itself is {\it odd} under time reversal. This implies, however, that
\begin{equation}
A^2 = -1.
\end{equation}
The existence of such an antiunitary symmetry is, by Kramers' theorem,\cite{gottfried} sufficient for the spectrum of $H_v$ to be at least doubly degenerate. We emphasize that this degeneracy has nothing to do with the electron spin, which is not included into $H_v$, nor with the true time reversal symmetry, which is a) broken by the mass terms in $H_v$, and b) even if it were not, it squares to $+1$, and by itself therefore does not imply any degeneracy.

 The uncovered Kramers' degeneracy of the spectrum of $H_v$ is independent of the particular representation. All four-dimensional representations  of the four matrices in $H_v$ are equivalent. In particular, one can always choose $\gamma_i$ to be real for $i=1,2$, and imaginary for $i=3,5$, \cite{herbut1} so that the pseudo time-reversal operator becomes easy to discern: $A= \gamma_3 \gamma_5  K$.

 \begin{table}[t]

\begin{tabular}{c | c c c c c  r }

State & mass & $\gamma_{35}$ & $I_t$ & A \\
 \hline
BLIS & $\gamma_0$ &+&+&+ \\
VFM$_1$& $\gamma_3$ &-&-&+ \\
VFM$_2$ & $\gamma_5 $&-&-&+ \\
QAH & $i\gamma_1 \gamma_2$ &+&-&-
\\

\end{tabular}

\caption[]{The properties of the broken layer-inversion symmetry (BLIS), two valley ferromagnets (VFM$_1$, VFM$_2$), and the quantum anomalous Hall state (QAH) of the spinless fermions in graphene bilayer under translations ($\gamma_{35}$), time reversal ($I_t$), and the pseudo time reversal symmetry ($A$).}
\end{table}

 {\it Zero modes.} -- The spectrum of $H_v$ contains two states with exactly zero energy. Assume that near the origin the amplitude vanishes linearly, $\Delta (r) = c r + O(r^2)$, so that the vortex configuration has a finite core energy. Discarding the higher order terms in this expansion, we can define the linearized vortex Hamiltonian
 \begin{equation}
 H_{v, lin} =   H_0 + c (x_1 \gamma_3  +  x_2 \gamma_5) ,
 \end{equation}
  so that $H_v = H_{v, lin} + O(r^2)$. Both $H_v$ and its linearized version $H_{v, lin}$ obey the pseudo time reversal symmetry, as well as the chiral symmetry:
  \begin{equation}
  \{H_v, \gamma_0 \} = \{ H_{v,lin},\gamma_0 \} = 0.
  \end{equation}
  Since $\gamma_0$ is even under pseudo time reversal, if the spectrum of $H_{v,lin}$ contains a Kramers' doublet at zero, both states will be the eigenstates of $\gamma_0$ with the {\it same} eigenvalue. One may then think of the original Hamiltonian $H_v$ as a continuous amplitude deformation of $H_{v,lin}$ which preserves the chiral and the pseudo time reversal symmetries, so that the zero modes cannot neither move nor split in the process.

  It therefore suffices to show that $H_{v, lin}$ has two zero modes. This is easiest to demonstrate in the momentum representation, in which $x_i = i\partial/\partial p_i$, and
 \begin{equation}
  H_{v, lin} = i\frac{\partial}{\partial p_1} \gamma_3 + i\frac{\partial}{\partial p_2} \gamma_5 + p^2 (  \gamma_2 \cos 2\theta_p + \gamma_1 \sin 2 \theta_p )
 \end{equation}
 where $(p, \theta_p)$ are the polar coordinates in the momentum space. In writing the last expression we have also introduced the dimensionless variables, so that the energy is measured in the units of $(c^2 /2 m )^{1/3}$, and the length in the units of $(2m c ) ^{-1/3}$.  The linearized Hamiltonian can now be viewed as describing a double vortex in {\it the momentum space}, and on rather general grounds one expects it to have {\it two} zero modes.\cite{jackiwrossi}  Indeed, they can be  analytically computed:
 \begin{equation}
 \Psi_1 = ( 0, e^{i \theta_p  } f(p), 0,   g(p))^T,
 \end{equation}
 \begin{equation}
 \Psi_2 = ( 0, g(p), 0,  e^{- i \theta_p   } f(p))^T,
 \end{equation}
 with
 \begin{equation}
 g(p) = {\cal N} p^2 K_{2/3} (p^3/3),
 \end{equation}
 with $K_{2/3} (z) $ as the modified Bessel function, $\cal{N}$ as a normalization factor, and $f(p)= -g' (p) / p^2$. Both $g(p)$ and $f(p)$ decrease exponentially at large momenta, and both are regular at small momenta. We also confirm that
 \begin{equation}
 \Psi_2 = A \Psi_1,
 \end{equation}
 by recalling that under time-reversal, $\theta_p \rightarrow \theta_{-p}=\theta_p + \pi$ as well. Both zero-modes have negative chirality: $\gamma_0 \Psi_i = -\Psi_i$. For an antivortex their chirality would be positive.

{\it Irreducible Skyrmion Hamiltonian.} -- The existence of two zero modes implies that when both states are occupied the electrical charge of the vortex is unity. This has interesting consequences for the charge of Skyrmion excitations. Before finally restoring spin and considering a spin-Skyrmion, let us define an auxiliary irreducible Hamiltonian
 \begin{equation}
 \tilde{H} (\vec{m} )  =  H_0 + m_1 (\vec{r}) \gamma_3   + m_2 (\vec{r}) \gamma_5 + m_3 (\vec{r}) \gamma_0,
\end{equation}
with the three-component vector $\vec{m}(\vec{r})$ describing a Skyrmion configuration: the mapping of the plane onto the two dimensional sphere $S_2$, at which the unit vector $\vec{m}(\vec{r})$ lives, with the Pontryagin index $ P = \int d^2 \vec{r} P(\vec{r})/4\pi =1$.
Here,
\begin{equation}
P (\vec{r}) =  \epsilon _{ijk} m_i \partial_x m_j \partial_y m_k = \sin\theta[(\partial_x\theta)( \partial_y \phi) - (\partial_y \theta )(\partial_x \phi)],
\end{equation}
and $\vec{m}(\vec{r})=(\sin\theta \cos\phi, \sin\theta\sin\phi, \cos\theta)$. By recognizing that
\begin{equation}
 P(\vec{r}) = [\cos\theta (\nabla\times \nabla\phi) - \nabla\times (\cos\theta \nabla \phi)]_z,
 \end{equation}
the Pontryagin index may be rewritten as
\begin{equation}
P = \frac{1}{2} [\sum_{\vec{r}_v} n(\vec{r}_v) \cos \theta(\vec{r}_v) - \cos\theta (R) \sum_{\vec{r}_v} n(\vec{r}_v) ],
\end{equation}
where,
\begin{equation}
n(\vec{r}_v)= \frac{1}{2\pi} \oint d\vec{l}\cdot \nabla \phi (\vec{r})
\end{equation}
is the vorticity at the point $\vec{r}_v$, encircled by the contour of integration. We assumed, for simplicity, that the angle $\theta$ is constant at the boundary $R$. The Skyrmion is usually depicted as a vortex with $n=1$ at the origin, and with the angle $\theta$ interpolating from $\theta(0)= 0$ to $\theta(R) = \pi$. In this case both the first and the second term in Eq. (15) contribute equally to $P$. But,
Eq. (15) implies that the Skyrmion is topologically equivalent to a particular meron-pair, consisting of a meron with $n=1$ (vortex) and $\theta=0$ at the origin, and of another meron with $n=-1$ (antivortex) and $\theta=\pi$  at some other point in space (Figure 1), when $P$ derives entirely from the first term. If we assume the two merons to be spatially well separated, simple counting leads to the elementary Skyrmion's total charge to be $2e$. Explicitly, the first meron at the origin is topologically equivalent to the configuration $\vec{m}= (x_1, x_2, -m)$, and the second meron at some distant position $\vec{L}$ is equivalent to $\vec{m}= (x_1-L_1 , - (x_2- L_2), +m)$. Since the zero modes of the vortex and the antivortex that make up the two merons have opposite chiralities, the charges of the two merons  are {\it the same} and equal to $e$. The charge of the irreducible Skyrmion is therefore $2e$.

\begin{figure}[t]
{\centering\resizebox*{80mm}{!}{\includegraphics{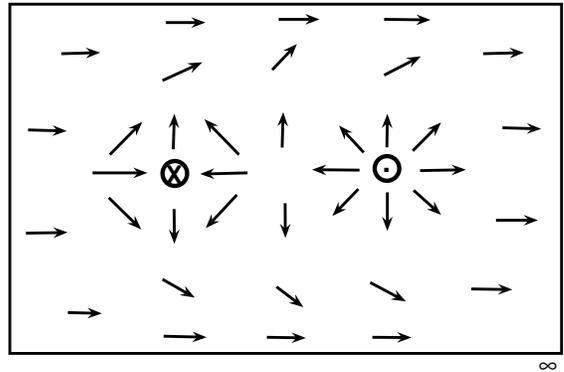}}
\par} \caption[] {The irreducible Skyrmion in the mass $\vec{m}$, as a pair of two merons, each one separately covering the northern and the southern hemisphere of the target space. Both merons carry a unit electric charge of the same sign.}
\end{figure}

{\it Charge of spin-Skyrmions.} -- All is ready now to restore the electron spin. Let us consider two prominent candidates for the ground state of the bilayer graphene: the QSH state,  and the LAF. Their representative masses are given by $\vec{N}\cdot \vec{\sigma} \otimes i\gamma_1 \gamma_2$ and $\vec{N}\cdot \vec{\sigma} \otimes \gamma_0$, with $\vec{N}$ as the vector order parameter,\cite{herbut3} respectively. Both states spontaneously break the rotational symmetry, and the latter also breaks the physical time reversal symmetry. Consider then a Skyrmion configuration for the order parameter $\vec{N}(\vec{r}) $:
\begin{equation}
 H (\vec{N} )  =  \sigma_0 \otimes H_0 + \vec{N}( \vec{r} ) \cdot \vec{\sigma} \otimes M,
\end{equation}
with either $M=i\gamma_1\gamma_2$ (QSH), or $M=\gamma_0$ (LAF). We will show that this Hamiltonian is {\it reducible}, and by some unitary transformation $U$ can be brought into a block-diagonal form:
\begin{equation}
U^\dagger H (\vec{N}) U = \tilde{H} (\vec{N}) \oplus \tilde{H} (\vec{N}'),
\end{equation}
with two possible results: either $\vec{N}' = +\vec{N}$, or $\vec{N}' = -\vec{N}$. The electrical charge of the spin-Skyrmion depends crucially on the outcome of this decomposition: the plus sign implies that the spin-Skyrmion is equivalent to the sum of the two Skyrmions in each irreducible block, so that the electrical charge of the two adds up to $4e$, whereas the negative sign implies the sum of a Skyrmion and an anti-Skyrmion, which leads to the charge zero.

The above result holds because  the five eight-dimensional matrices $\sigma_0 \otimes \gamma_1$, $\sigma_0 \otimes \gamma_2$ and $\vec{\sigma}\otimes M$ featured in $H(\vec{N})$ all mutually anticommute and square to unity, and as such form a representation of the Clifford algebra $C(5,0)$. \cite{ryu, herbut1} The crucial fact about $C(5,0)$ is that it has {\it two} inequivalent irreducible representations, which are both four dimensional, and differ in the sign of an odd number of matrices,\cite{remark} which we here take to be three. The plus (minus) sign in the above decomposition of $H(\vec{N})$ then corresponds to two equivalent (inequivalent) representations of $C(5,0)$ appearing in the two blocks.

To determine the charges of the two spin-Skyrmions under study one therefore in principle needs to block-diagonalize their corresponding Hamiltonians in Eq. (17). The full calculation can be avoided, however, by simply checking the value of the trace of the product of the five matrices in $H(\vec{N})$
\begin{equation}
Q= \frac{1}{2} Tr \prod _{i=1,2} \sigma_0 \otimes \gamma_i \prod_{k=1,2,3} \sigma_k \otimes M.
\end{equation}
 Since $Q=4$  for $M=i\gamma_1 \gamma_2$ and $Q=0$  for $M=\gamma_0$, it may serve to distinguish between  the two spin-Skyrmions. Being by construction invariant under any unitary transformation, $Q$  may also be evaluated from the block-diagonal forms in Eq. (18):
\begin{equation}
Q= \frac{1}{2} Tr (\sigma_0)^2 (\sigma_k)^3 \otimes \gamma_1 \gamma_2 \gamma_3 \gamma_5 \gamma_0,
\end{equation}
and $Q=4$ for $\vec{N}=\vec{N}'$ (when $k=0$), and $Q=0$ for $\vec{N}=-\vec{N}'$ (when $k=3$). So, $\vec{N}=+ \vec{N}'$ ($\vec{N}=-\vec{N}'$) corresponds to the QAH (LAF) state, and $Q$ may be identified with the electrical charge of the spin-Skyrmion.

{\it Topological superconductivity.} -- Condensation of charged spin-Skyrmions
would produce a superconductor with the elementary charge of $4e$, as discussed by Fr$\ddot{o}$hlich in the pre-BCS era. Fr$\ddot{o}$hlich's mechanism for ideal conductivity in one dimension has been extended to and greatly elaborated in higher dimensions by Wiegmann \cite{pasha}, and discussed in the context of cuprates \cite{dhlee} and, recently, chalogenides\cite{baskaran}. The novel point in our example is that due to the quadratic band crossing the charge of the Skyrmion becomes doubled, which makes the topological mechanism for superconductivity more efficient, and the contrast with the BCS paradigm even more striking.

{\it Conclusion and summary.} -- We demonstrated that the charge of Skyrmion textures in a graphene bilayer with quadratic band crossing is doubled relative to its single layer equivalent. The basis for our result is the explicit derivation of two zero modes for the basic vortex Hamiltonian with quadratic dispersion, and the identification of the hidden pseudo time reversal symmetry which protects their degeneracy. It was argued that the condensation of the Skyrmion textures of the QSH state in bilayer graphene upon doping leads to a non-BCS superconductor with the flux quantized in units  $hc/4e$.

 The present derivation of the charge of the Skyrmion in the QSH state can be applied equally well to the single layer graphene Hamiltonian, in which case it yields the charge of $2e$, in agreement with \cite{grover}. Our approach circumvents the use of the gradient expansion and of the concomitant Wess-Zumino-Witten term \cite{abanov}, the applicability of which to systems with the quadratic band crossing does not seem obvious\cite{yao}, but has recently shown to be possible. \cite{moon}

The  nature of the doubling discussed here is different from the one for the bilayer graphene in the magnetic field,\cite{ed} or in one-dimensional systems with quadratic band crossing \cite{yao}, when only the number of zero modes, and not the whole spectrum, is doubled, and the pseudo time reversal symmetry introduced here is absent.

This work was supported by the NSERC of Canada and in part by the National Science Foundation under Grant No. NSF PHY05-51164. Many useful discussions and the collaboration in the early stages of this project with B. Roy are gratefully acknowledged.


\begin{thebibliography}{99}

\bibitem{sun} K. Sun {\sl et al.}, Phys. Rev. Lett. {\bf 103}, 046811 (2009).

\bibitem{velasco} J. Velasco Jr. {\sl et al.}, Nat. Nano. {\bf 7}, 156 (2012).
\bibitem{feldman} B. E. Feldman, J. Martin, and A.  Yacoby, Nat. Phys. {\bf 5}, 889 (2009).
\bibitem{weitz} R. T. Weitz {\sl et al.},  Science {\bf 330}, 812 (2010).
\bibitem{freitag} F. Freitag {\sl et al.}, Phys. Rev. Lett. {\bf 108}, 076602   (2012).
\bibitem{mayorov} A. S. Mayorov {\sl et al.},  Science {\bf 333}, 860 (2011).

\bibitem{nandkishore} R. Nandkishore, L.  Levitov, Phys. Rev. B {\bf 82}, 115124 (2010).
\bibitem{zhang} F. Zhang {\sl et al.}, Phys. Rev. Lett. 106, 156801 (2011).
\bibitem{jung} J. Jung {\sl et al.},  Phys. Rev. B 83, 115408 (2011).
\bibitem{zhang1} F. Zhang {\sl et al.}, Phys. Rev. B 81, 041402 (R) (2010).
\bibitem{lemonik} Y. Lemonik {\sl et al.}, Phys. Rev. B 82, 201408 (2010).
\bibitem{vafekyang} O. Vafek and K. Yang, Phys. Rev. B 81, 041401 (2010).
\bibitem{vafek} O. Vafek, Phys. Rev. B {\bf 82}, 205106 (2010).
\bibitem{uebelacker} S. Uebelacker and C. Honerkamp, Phys. Rev. B {\bf 84}, 205122 (2011).

\bibitem{hou} C-Y. Hou, C. Chamon, and C. Mudry, Phys. Rev. Lett. {\bf 98}, 186809 (2007).
\bibitem{ryu} S. Ryu, {\sl et al.}, Phys. Rev. B {\bf  80}, 205319   (2009).
\bibitem{herbut1} I. F. Herbut, Phys. Rev. Lett. {\bf  104}, 066404   (2010);
Phys. Rev. B {\bf 85}, 085304 (2012), and references therein.
\bibitem{jackiwrossi} R. Jackiw and P. Rossi, Nucl. Phys. B {\bf 190}, 681 (1981).

\bibitem{herbut2} I. F. Herbut, Phys. Rev. Lett. {\bf 97}, 146401 (2006).

\bibitem{pasha} P. B. Wiegmann, Phys. Rev. B {\bf 59}, 15705 (1999).
\bibitem{ed} E. McCann and V. I. Fal'ko, Phys. Rev. Lett. {\bf 96}, 086805 (2006).

\bibitem{herbut'}I. F. Herbut, V. Juri\v ci\' c, B. Roy,  Phys. Rev. B {\bf 79}, 085116 (2009).



\bibitem{dhlee} D.-H. Lee, Phys. Rev. B {\bf 60}, 12429 (1999).
\bibitem{baskaran} G. Baskaran, arXiv:1108.3562.

\bibitem{gottfried} K. Gotfried and T-M. Yan, {\sl Quantum Mechanics: Fundamentals}, 2nd ed., (Springer, 2004).
\bibitem{herbut3} I. F. Herbut, Phys. Rev. Lett. {\bf 99}, 206404 (2007).

\bibitem{remark} For example, $\{\gamma_0, \gamma_1,\gamma_2, \gamma_3, \gamma_5 \}$ and $\{\gamma_0, \gamma_1,-\gamma_2, -\gamma_3, -\gamma_5 \}$.
 No four dimensional matrix commutes with two while anticommuting with three of $\gamma$-matrices, so the representations are inequivalent.

\bibitem{grover} T. Grover and T. Senthil, Phys. Rev. Lett. {\bf 100}, 156804 (2008).

\bibitem{abanov} A. G. Abanov and P. B. Wiegmann, Nucl. Phys. B, {\bf 570}, 685 (2000).
\bibitem{yao} H. Yao and D.-H. Lee, Phys. Rev. B {\bf 82}, 245117 (2010).
\bibitem{moon} E.-G. Moon, Phys. Rev. B {\bf 85}, 245123 (2012).
\end{thebibliography}
\end{document}